\documentclass[aps,10pt,prd,tightenlines,twocolumn,twoside,showpacs,nofootinbib,superscriptaddress]{revtex4-2}
\usepackage[utf8x]{inputenc}
\usepackage[T1]{fontenc}
\usepackage{alltt}
\usepackage{amsmath}
\usepackage{amsthm}
\usepackage{amssymb}
\usepackage{bm}
\usepackage{booktabs,siunitx}
\usepackage{cancel}
\usepackage{color}
\usepackage{csquotes}
\usepackage{enumerate}
\usepackage[marginal]{footmisc}

\usepackage{graphicx,bm}
\usepackage{lipsum}
\usepackage{listings}
\usepackage{mathtools}
\usepackage{mathrsfs}
\usepackage{physics}
\usepackage{ragged2e}
\usepackage{relsize}
\usepackage{slashed}
\usepackage{soul}
\usepackage{spverbatim}
\usepackage{wrapfig}
\usepackage{xcolor}
\usepackage{diagbox}
\usepackage{multirow}
\usepackage{subcaption}

\usepackage[pdftex,hypertexnames=false,linktocpage=true]{hyperref}
\hypersetup{colorlinks=true,linkcolor=blue,anchorcolor=blue,citecolor=darkgreen,filecolor=blue,urlcolor=blue,bookmarksnumbered=true,
	pdfview=FitB
}
\colorlet{darkgreen}{green!60!black}
\colorlet{brightyellow}{yellow!75!red}
\colorlet{orange}{red!50!yellow}
\colorlet{darkblue}{blue!60!black}
\colorlet{darkred}{red!80!black}
\colorlet{greenblue}{green!50!blue}

\makeatletter

\newcommand{\Rmnum}[1]{\expandafter\@slowromancap\romannumeral #1@}
\makeatother

\def\dd{{\mathrm{d}}}
\def\imag{{\mathrm{i}}}

\begin{document}
%\begin{CJK}{UTF8}{gbsn}
\title{Electron form factors in Basis Light-front Quantization}

\author{Lingdi Meng}
\email{lingdimeng@foxmail.com}

\affiliation{%
	Department of Physics, Hebei University, Baoding, 071002, China
}%
\affiliation{%
	Institute of Modern Physics, Chinese Academy of Sciences, Lanzhou, Gansu 730000, China
}%

\author{Shuo Tang}
\email{tang@iastate.edu}
\affiliation{%
	Department of Physics and Astronomy, Iowa State University, Ames, Iowa 50011, USA
}%

\author{Zhi Hu}
\email[corresponding author: ]{huzhi0826@gmail.com}
\affiliation{%
	High Energy Accelerator Research Organization (KEK), Ibaraki 305-0801, Japan
}%

\author{Guo-Li Wang}
\email{wgl@hbu.edu.cn}
\affiliation{%
	Department of Physics, Hebei University, Baoding, 071002, China
}%
\affiliation{%
	Key Laboratory of High-precision Computation and Application of Quantum Field Theory of Hebei Province, Baoding, China
}%

\author{Yang Li}
\email{leeyoung1987@ustc.edu.cn}
\affiliation{%
Department of Physics and Astronomy, Iowa State University, Ames, Iowa 50011, USA
}	
\affiliation{%
Department of Modern Physics, University of Science and Technology of China, Hefei 230026, China
}

\author{Xingbo Zhao}
\email{xbzhao@impcas.ac.cn}
\affiliation{%
	Institute of Modern Physics, Chinese Academy of Sciences, Lanzhou, Gansu 730000, China
}
\affiliation{%
	School of Nuclear Science and Technology, University of Chinese Academy of Sciences, Beijing 100049, China
}

\author{James P. Vary}
\email{jvary@iastate.edu}
\affiliation{%
	Department of Physics and Astronomy, Iowa State University, Ames, Iowa 50011, USA
}

\date{\today}% It is always \today, today,
\footnote{First Author and Second Author contribute equally to this work.}

\begin{abstract}
	
In this paper, we evaluate the electromagnetic and gravitational form factors as well as the corresponding generalized parton distributions of the electron using the Basis Light-front Quantization approach to QED. We compare our results with those from light-front perturbation theory. We adopt a novel basis with its scale depending on the constituents' longitudinal momentum fraction. We show that this basis improves convergence of the form factors with increasing basis dimension, compared to that calculated in the original basis with fixed scale. These results both validate the BLFQ approach and provide guidance for its efficient implementation in solving light-front Hamiltonian mass eigenstates for more complex systems in QED and QCD.

%We are able to obtain reasonably accurate result with perturbation theory by extrapolating to complete basis limit.

\end{abstract}

\maketitle

\section{Introduction}

Describing the structure of relativistic bound states is one of the fundamental challenges of nuclear and hadronic physics. Among various approaches, Basis Light-front Quantization (BLFQ) has emerged as a promising framework to solve for the nonperturbative dynamics of quantum chromodynamics (QCD)~\cite{PhysRevC.81.035205}.  The applications of BLFQ in quantum electrodynamics (QED)~\cite{honkanen2011,PhysRevD.89.116004,ZHAO201465,PhysRevD.91.105009,huzhi2021,sreeraj2022,sreeraj2023} and QCD~\cite{Li:2013cga,PhysRevD.96.016022,LI2016118,PhysRevD.98.114038,PhysRevC.99.035206,lanjiangshan2019m,xusiqi2021b,adhikari2021m,lanjiangshan2022m,liuyiping2022b,kuangzhongkui2022t,pengtiancai2022b,huzhi2022b,zhuzhimin2023m,zhuzhimin2023m1,xusiqi2023b,linbolang2023b,gross2023} have shown considerable success. This approach takes advantage of the light-front dynamics and the Hamiltonian formalism, offering intuitive insights into bound state structure. Nevertheless, further efforts are needed to both validate the approach with applications solvable by other means as well as to provide avenues for improved efficiency. These are the dual goals of the present work.

In recent years, the study of gravitational form factors of hadrons has received renewed interest~\cite{PhysRevD.99.014511,doi:10.1142/S0217751X18300259,Burkert2018,PhysRevD.98.034009}. Even though it is impractical to measure the gravitational form factors directly through the coupling of hadrons with a graviton, it is possible to determine them via the corresponding generalized parton distributions (GPDs) which are measured through deeply virtual Compton scattering~\cite{PhysRevD.55.7114,DIEHL200341,GOEKE2001401}. Such a measurement is one of the key goals of the upcoming electron-ion colliders~\cite{Accardi2016,Kalantarians_2014}.

As support for applying BLFQ to problems in QCD, we investigate the electron system in QED. The structure of the physical electron serves as a benchmark for GPDs~\cite{PhysRevD.81.013002,PhysRevD.89.116004,PhysRevD.90.113001}, transverse momentum distributions (TMDs)~\cite{Mantovani:2016kip} and spin decomposition~\cite{PhysRevD.79.071501,PhysRevD.91.017501,PhysRevD.93.054013}, etc, for bound states in QCD. Therefore, we study the electromagnetic and gravitational form factors and their corresponding GPDs of the electron with BLFQ, and compare our results with the results from the light-front perturbation theory~\cite{BRODSKY2001311}. In addition, we employ a novel basis where the transverse harmonic oscillator basis parameter, $b$, is $x$-dependent, and compare the results with the original fixed $b$ basis and demonstrate improved convergence.

The paper is organized as follows.
We begin with the electromagnetic and gravitational form factors and GPDs within the overlap representation in Sec. \ref{sec2}. Then we introduce the theoretical framework, including the choice of naive basis and the corresponding truncation, as well as the $x$-dependent $b$ basis and compare the resulting form factors with those evaluated in the fixed $b$ basis in Sec. \ref{sec3}. Then in Sec. \ref{sec4}, we present the numerical results and compare them with light-front perturbation theory. Finally we conclude in Sec. \ref{sec5}.

\section{\label{sec2}Form Factors and GPDs}
In this paper, we focus on the physical electron in QED as a test of BLFQ. The physical electron can be expanded in the Fock space as:
\begin{equation}
\label{fexpand}
\ket{e_{\text{phys}}} = c_1\ket{e} +c_2\ket{e\gamma} + c_3\ket{e\gamma\gamma} +  c_4\ket{ee\bar{e}} + ...
\end{equation}
where $c_n$ schematically denotes the coefficient for basis states in each Fock sector. The initial and the final momenta of the electron system in the frame we choose are
\begin{equation}
P=(P^+, 0_\perp, \frac{M^2}{P^+}),
\end{equation}
\begin{equation}
P'=((1-\zeta)P^+,-\Delta_{\perp}, \frac{M^2+\Delta_{\perp}^2}{(1-\zeta) P^+}),
\end{equation}
respectively, and the momentum transfer is
\begin{equation}
\Delta=P-P^{\prime}=(\zeta P^+, \Delta_{\perp}, \frac{t+\Delta_{\perp}^2}{\zeta P^+}),
\end{equation}
where $t=\Delta^2$, and $M$ is the mass of the system. For $\zeta=0$, $t=-\Delta_{\perp}^2$.

The light-cone Fock expansion of the state $\ket{\Psi}$ with the momentum $P$ and the helicity $\lambda$ is
\begin{equation}
\begin{aligned}
 &\ket{\Psi^\lambda (P)}\\
 = 
 &\sum_n \sum_{\lambda_1 \ldots \lambda_n} 
 \int \prod_{i=1}^n[\frac{d x_i d^2 k_{\perp i}}{\sqrt{x_i} 16 \pi^3}] 16 \pi^3 \\
 &\times\delta(1-\sum_{i=1}^n x_i) 
 \delta^2 (\sum_{i=1}^n \vec{k}_{\perp i})
 \Psi_{\lambda_1 ... \lambda_n}^\lambda (\{x_i, k_{\perp i} \}) \\
 &\times\ket{\{n, x_i P^+, x_i P_\perp+{k_{\perp i}},\lambda_i \}}.
\label{iexpand}
\end{aligned}
\end{equation}
Here $x_i=p_i ^+ /P^+$ is the longitudinal momentum fraction, $k_{\perp i}$ represents the relative transverse momentum and $\lambda_i$ labels the light-cone helicity for the $i$-th constituent within the Fock sector. The physical transverse momentum is $p_{\perp i}=x_i P_\perp + k_{\perp i}$. $n$ is the number of particles in a Fock sector. $\Psi_{\lambda_1 ... \lambda_n}^\lambda (\{x_i, \vec{k}_{\perp i} \})$ are the boost invariant light-front wave functions (LFWFs).

First, we study the electromagnetic form factors $F_1(t)$ and $F_2(t)$, which are known as the Dirac form factor and the Pauli form factor, respectively. In the case of a spin$-\frac{1}{2}$ composite system, they are defined by:
\begin{equation}
\begin{aligned}
&\mel{\Psi^{\lambda'} (P')}{J^\mu(0)}{\Psi^\lambda (P)}\\
=&\bar{u}(P', \lambda') \Big[F_1(t) \gamma^\mu +F_2(t) \frac{\imag}{2M} \sigma^{\mu\alpha}(-\Delta_\alpha) \Big]u(P, \lambda),
\end{aligned}
\end{equation}

where $u(P, \lambda)$ is the bound state spinor, and $M$ is the mass of the bound state. 

We adopt the ``good current'' $J^+$ which identifies the Dirac and the Pauli form factors from its helicity-conserving and helicity-flip vector current matrix elements respectively. Considering the boost invariance of the LFWFs, the state with nonzero transverse momentum can be related to that with zero transverse momentum, so the electromagnetic form factors of the physical electron with the mass $M_e$ can be separated as 
\begin{equation}
\begin{aligned}
\mel{e^\uparrow_\text{phys}(\Delta_\perp)}{\frac{J^+(0)}{2P^+}}{e^\uparrow_\text{phys}(0_\perp) } & = F_1(t),\\
\mel{e^\uparrow_\text{phys}(\Delta_\perp) }{ \frac{J^+(0)}{2P^+}}{e^\downarrow_\text{phys}(0_\perp) } & = -(\Delta_1 -\imag \Delta_2) \frac{F_2(t)}{2M_e}.
\end{aligned}
\end{equation}
Here the state vector $\ket{e^{\uparrow(\downarrow)}_\text{phys}(\Delta_\perp)}$ stands for a physical electron state with helicity (anti-) parallel to the direction of $P^+$ and the center of mass transverse momentum $\Delta_\perp$. 
One can represent the helicity-flip matrix elements using the helicity-conserving matrix elements by exploiting the transverse parity symmetry~\cite{BRODSKY1998299,PhysRevD.73.036007}.

The gravitational form factors for a spin$-\frac{1}{2}$ composite system are defined by the energy-momentum tensor~\cite{PhysRevLett.78.610}. They can be written in terms of fermion (f) and boson (b) contributions separately as,
\begin{equation}
\begin{aligned}
&\mel{\Psi^{\lambda'} (P')}{T^{\mu \nu}_\text{f,b}(0) }{\Psi^{\lambda} (P)} \\
=&\bar{u} (P', \lambda') \Big[ A_\text{f,b}(t) \gamma^{(\mu}\overline{P}^{\nu)}
+B_\text{f,b}(t) \frac{\imag}{2M}\overline{P}^{(\mu} \sigma^{\nu)\alpha}(-\Delta_\alpha)\\
&+ C_\text{f,b}(t)\frac{1}{M}(\Delta^\mu \Delta^\nu - g^{\mu \nu}\Delta^2) 
+ \bar{C}_\text{f,b} (t) g^{\mu \nu}M
\Big]u(P, \lambda),
\end{aligned}
\label{eq13}
\end{equation}
where $\overline{P}^\mu = (P'^\mu+P^\mu)/2$, $a^{(\mu}b^{\nu)} = (a^\mu b^\nu + a^\nu b^\mu)/2$. Since we are only interested in $A_\text{f,b}(t)$ and $B_\text{f,b}(t)$, we choose the Drell-Yan frame and only consider the non-interacting parts of the energy-momentum tensor $T^{++}(0)$. Analogous to $F_1(t)$ and $F_2(t)$, we obtain the gravitational form factors of the physical electron:
\begin{equation} 
\begin{aligned}
\mel{e^\uparrow_\text{phys}(\Delta_\perp) }{\frac{T^{++}_\text{f,b}(0)}{2(P^+)^2} }{ e^\uparrow_\text{phys}(0_\perp)} & = A_\text{f,b}(t), \\
\mel{e^\uparrow_\text{phys}(\Delta_\perp)}{\frac{T^{++}_\text{f,b}(0)}{2(P^+)^2}}{e^\downarrow_\text{phys}(0_\perp) }& =
-(\Delta_1 -   \imag \Delta_2)  \frac{B_\text{f,b}(t)}{2M_e}.
\end{aligned}
\end{equation}
By choosing the $\mu\nu=++$  component, the terms associated with form factor $C_\text{f,b}(t)$ and $\bar{C}_\text{f,b} (t)$ in Eq.~\eqref{eq13} vanish.
$A(t) = A_\text{f}(t)+A_\text{b}(t)$ and $B(t) = B_\text{f}(t)+B_\text{b}(t)$ are the total gravitational form factors which consist of the contribution from the constituent fermion and boson. Note that these form factors receive the contribution from all the Fock sectors, which in our current truncation scheme includes only the $\ket{e}$ and $\ket{e\gamma}$ sectors.

In order to provide a more detailed view of the charge and matter distribution inside the physical electron, we also investigate the GPDs of the electron, which are universal nonperturbative objects used to describe hard exclusive processes.

$H_\text{f,b}$ and $E_\text{f,b}$ are the leading twist (twist-2) off-forward parton distributions defined for fermion and boson, respectively~\cite{BRODSKY200199,DIEHL200341,PhysRevD.76.034002}:
\begin{equation}
\begin{aligned}
&\int \frac{\dd z^-}{8\pi}\exp(\imag x P^+ z^- /2) \\
&\times\mel{\Psi^{\lambda'} (P')}{\overline{\psi}(0) \gamma^+ \psi(z)}{\Psi^{\lambda} (P)} \Big|_{z^+=0, z^\perp = 0} \\
=&\frac{1}{2P^+}  \bar{u}(\bm{P}',\lambda')  \Big[
H_\text{f}(x,\zeta,t) \gamma^+\\
&+E_\text{f}(x,\zeta,t) \frac{\imag \sigma^{+j}(-\Delta_j)}{2M}
\Big]u(\bm{P},\lambda),
\end{aligned}
\end{equation}
and
\begin{equation}
\label{eq_GPDphoton}
\begin{aligned}
&\frac{1}{2P^+}\int \frac{\dd z^-}{2\pi}\exp(\imag x P^+ z^- /2)\\
&\times \mel{\Psi^{\lambda'} (P')}{F^{+\mu}(0) F_\mu^+(z)}{\Psi^{\lambda} (P)} \Big|_{z^+=0, z^\perp = 0} \\
=&\frac{1}{2 P^+}  \bar{u}(\bm{P}',\lambda')  \Big[
H_\text{b}(x,\zeta,t) \gamma^+\\
&+E_\text{b}(x,\zeta,t)  \frac{\imag \sigma^{+j}(-\Delta_j)}{2M}
\Big]u(\bm{P},\lambda).
\end{aligned}
\end{equation}

From the definition, $H$ is associated with the helicity-conserving amplitude, while $E$ is with the helicity-flipping amplitude.
In this work, we concentrate only on the zero skewness limit $\zeta=0$. In this limit, only the diagonal process ($n{\rightarrow}n$) contributes to the GPDs.

Thus in the overlap representations, the GPDs for the constituents within a composite are given by
\begin{equation}
\begin{aligned}
&H^q(x, 0, t)\\
= & \sum_{n, \lambda_i} \int \prod_{i=1}^n \frac{d x_i d^2 k_{\perp i}}{16 \pi^3} 16 \pi^3 \delta\left(1-\sum_j x_j\right) \\
& \times \delta^2\left(\sum_{j=1}^n k_{\perp j}\right) \delta\left(x-x_q\right) \\
& \times \Psi_{\lambda_1 ... \lambda_n}^{\uparrow *} (\{x_i ', k_{\perp i} '\}) \Psi_{\lambda_1 ... \lambda_n}^\uparrow (\{x_i, k_{\perp i} \}) ,
\end{aligned}
\end{equation}
and
\begin{equation}
\begin{aligned}
&\frac{\Delta^1-i \Delta^2}{2 M} E^q(x, 0, t)\\
= & \sum_{n, \lambda_i} \int \prod_{i=1}^n \frac{d x_i d^2 k_{\perp i}}{16 \pi^3} 16 \pi^3 \\
& \times \delta\left(1-\sum_j x_j\right) \delta^2\left(\sum_{j=1}^n k_{\perp j}\right) \delta\left(x-x_q\right) \\
& \times \Psi_{\lambda_1 ... \lambda_n}^{\uparrow *} (\{x_i ', k_{\perp i} '\}) \Psi_{\lambda_1 ... \lambda_n}^\downarrow (\{x_i, k_{\perp i} \}),
\end{aligned}
\end{equation}
 where $q$ labels the struck parton and $x'_q =x_q$, $k'_{\perp q} = k_{\perp q} -(1-x_q)\Delta_\perp$; for the other partons (i.e. spectators) $x'_i =x_i$, $k'_{\perp i} = k_{\perp i} + x_i\Delta_\perp$ ($i = 1, ..., n$ with $q$ excluded). $\Psi_{\lambda_1 ... \lambda_n}^{\lambda}$ are the LFWFs in Eq.~\eqref{iexpand}, and can be written in terms of the harmonic oscillator basis functions, see Sec. \ref{sec3}. For all these observables, we also obtain the perturbation theory results for comparison, see Sec. \ref{sec4}.

In addition, from the first moment of the GPDs at zero skewness, one gets the following sum rules for $F_1$ and $F_2$~\cite{PhysRevLett.78.610},
\begin{equation}
\begin{aligned}
F_{1\ \text{f,b}}(t)=\int_0^1 H_\text{f,b}(x,\zeta=0,t) \dd x,\\ 
F_{2 \ \text{f,b}}(t)=\int_0^1 E_\text{f,b}(x,\zeta=0,t)  \dd x;
\end{aligned}
\end{equation}
and the gravitational form factors are related to the second moment of the GPDs,
\begin{equation}
\begin{aligned}
A_\text{f,b}(t)=\int_0^1 x H_\text{f,b}(x,\zeta=0,t)  \dd x,\\
B_\text{f,b}(t)=\int_0^1 x E_\text{f,b}(x,\zeta=0,t)  \dd x.
\end{aligned}
\end{equation}

\section{\label{sec3}Basis Light-Front Quantization}

BLFQ is based on the light-front Hamiltonian formalism~\cite{PhysRevC.81.035205}, and it aims to solve the light-front eigenvalue equation
\begin{equation}
P^\mu P_\mu \ket{\Psi} = M^2 \ket{\Psi},
\label{eq01}
\end{equation} 
 
where $P^-$ is the light-front Hamiltonian, and $M^2$ is the squared invariant mass of the bound state.
The eigenvector $\ket{\Psi}$ can be expanded with Fock space representation as  Eq.~\eqref{iexpand}. 
With the eigenvector $\ket{\Psi}$, one can investigate the observable of interest by computing its matrix element: $\expval{\mathcal{O}} = \mel{\Psi}{\mathcal{O}}{\Psi}$.
In this work, we truncate the Fock sectors in Eq.~\eqref{fexpand} of a physical electron up to the first two sectors, namely 
\begin{equation}
\label{focktrunc}
\ket{e_\text{phys}} =c_1 \ket{e} + c_2\ket{e\gamma}.
\end{equation}
The light-front QED Hamiltonian $P^-$ within the light-cone gauge (i.e. $A^+=0$) takes the following form, omitting the instantaneous-photon/fermion interaction since it should be included only when the two-photon Fock sector is added~\cite{ZHAO201465}:  
\begin{equation}
\begin{aligned}
P^-=&\int \dd^2 \vec{x}^\perp \dd x^- \Big[\frac{1}{2} \overline{\psi} \gamma^+ \frac{m^2_e +(\imag\partial^\perp)^2}{\imag\partial^+}\psi \\
&+\frac{1}{2}A^j (\imag \partial^\perp)^2 A^j +ej^\mu A_\mu   \Big],
\label{hamiltonian}
\end{aligned}
\end{equation}

here $\psi$ and $A_\mu$ are the field operators of the fermion and gauge boson, respectively; $m_e$ and $e$ are the mass and charge of a bare electron.

In BLFQ, the basis we adopt is discretized. For each constituent, its longitudinal motion is described by a plane wave $e^{-\text{i}p^+ x^- /2}$, which is confined inside a box of length $2L$; then we impose the (anti-) periodic boundary conditions for (fermions) bosons. Thus the longitudinal momentum fraction of the $i$-th constituent is discretized as $x_i = p_i^+/P^+=k_i/K_{\text{tot}}$, with $k_i$ being (half-) integers for (fermions) bosons. Note that for bosons, $k_i$ takes values from 1 instead of 0 since we omit the zero mode. Thus, $K_{\text{tot}}$ is our longitudinal truncation parameter.

The transverse motion is represented by a 2-dimensional (2D) harmonic oscillator (HO) basis function, 
\begin{equation}
\begin{aligned}
\phi_{nm}({\vec{p}_\perp})=&\frac{1}{b}\sqrt{\frac{4\pi n!}{(n+\abs{m})!}}\Big(\frac{p_\perp}{b}\Big)^{\abs{m}}\\ 
&\times e^{-\frac{1}{2} p_\perp ^2/b^2} L^{|m|}_n (p_\perp ^2/b^2) e^{\imag m\theta_p},
\end{aligned}
\label{hobasis}
\end{equation}
where $n$ and $m$ are the principal and angular quantum numbers, respectively. $b$ sets the momentum scale of the basis,
 
$L_n^{\abs{m}}$ is the associated Laguerre polynomial, and $\theta_p = \arg \vec{p}_\perp$.
We truncate the infinite basis in transverse directions by the truncation parameter $N_\text{max}$,
so that the retained basis states satisfy
\begin{equation}
\sum_i 2n_i +\abs{m_i}+1 \le N_\text{max},
\label{sum}
\end{equation} 
where the summation runs over all the constituents.
Notice that enlarging $N_\text{max}$ would not only increase the resolution but also broaden the infrared and ultraviolet coverage for the transverse momenta, viz. $ \lambda_\text{IR}\sim b/\sqrt{N_\text{max}}$  and $  \Lambda_\text{UV}\sim b \sqrt{ N_\text{max}}$.
Thus $\bar \beta \equiv \{k, n, m, \lambda\}$ denotes the complete single-particle quanta.

With the formalism given by Eq.~\eqref{eq01} to Eq.~\eqref{sum}, one can write down the matrix of the light-front QED Hamiltonian in the basis representation with regularization achieved at the scales defined by $N_\text{max}$, $b$, and $K_\text{tot}$. Upon diagonalization of the Hamiltonian, the lowest mass eigenstate is identified as the physical electron state, $\ket{e_\text{phys}}$~\cite{Zhao2012,PhysRevD.91.105009}. Then for the first and the second Fock sectors of $\ket{e_\text{phys}}$, the LFWFs can be written as 
\begin{equation}
\begin{aligned}
\Psi_{\lambda_e}^\lambda (x_e, \vec{k}_{\perp e}) = \sum_{{n_e, m_e}} [\psi^\lambda {(\bar\beta_e)}
  \phi_{n_e m_e}({\vec{k}_{\perp e}})],
\end{aligned}
\end{equation}
\begin{equation}
\begin{aligned}
&\Psi_{\lambda_e, \lambda_\gamma}^\lambda (x_e, \vec{k}_{\perp e}, x_{\gamma}, \vec{k}_{\perp \gamma }) \\
=& \sum_{{n_e, m_e, n_{\gamma}, m_{\gamma}}} [\psi^\lambda {(\bar\beta_e, \bar\beta_{\gamma} )}
  \phi_{n_e m_e}({\vec{k}_{\perp e}}) \phi_{n_{\gamma} m_{\gamma}}({\vec{k}_{\perp {\gamma}}})],
\end{aligned}
\end{equation}
respectively, where $\psi^\lambda {(\bar\beta_e)} = \bra{\bar\beta_e}\ket{e_\text{phys}}$ and $\psi^\lambda {(\bar\beta_e, \bar\beta_{\gamma} )} = \bra{\bar\beta_e, \bar\beta_{\gamma}}\ket{e_\text{phys}}$ are the components of the eigenvectors obtained by diagonalizing the Hamitonian in Eq.~\eqref{eq01}. 
With the LFWFs, we can calculate the observables.

\subsection{The $x$-dependent $b$ Basis}
In the previous calculations of the electron~\cite{PhysRevD.89.116004,PhysRevD.103.036005}, the scale parameter $b$ in the HO basis function was a constant over $x$.  However in this work, we introduce an $x$-dependent $b$ basis with $b$ in the HO depending on the longitudinal momentum fraction of the $i$-th particle, i.e. $b_i=b' \sqrt{x_i}$, where $b'$ is an $x$-independent dimensionful constant. This $x$-dependent $b$ basis is matched with the fixed $b$ HO basis (i.e. $x$-independent $b$ basis) where the momentum $\vec{q}_{i\perp}$ is associated with our momentum $\vec{p}_{i\perp}$ by $\vec{q}_{i\perp} = \vec{p}_{ i \perp}/\sqrt{x_i} $~\cite{maris2013bound,PhysRevD.91.105009,Li:2013cga}.

This choice of basis allows for the exact factorization between the center-of-mass (c.m.) motion and the intrinsic motion~\cite{PhysRevD.91.105009}. For two-body systems, one can easily get the form of relative (Jacobi) coordinates to separate the wave function into “intrinsic” and “c.m.” components, with which both choices of HO basis work. However for systems with more constituents, the $x$-independent $b$ basis loses its efficacy to extract the c.m. motion and the intrinsic motion. 
We take a proton as an example, whose first Fock sector has 3 constituents. According to Talmi-Moshinsky (TM) transformation~\cite{MOSHINSKY1959104,YOUPING1985387}, with the $x$-dependent $b$ basis, we get three Jacobi momenta by choosing TM transformation angles $\delta$ as $\tan\delta_1 = \sqrt{x_3/x_1}$ and $\tan\delta_2 = \sqrt{(x_2 + x_3)/x_1}$:
\begin{equation}
\begin{aligned}
&P = p_1 + p_2 +p_3,\\
&l_1 = p_1(x_2+x_3) - (p_2+p_3)x_1,\\
&l_2 = \frac{x_3p_2 - x_2p_3}{x_2+x_3},
\label{Jacobi}
\end{aligned}
\end{equation}
 
while with the fixed $b$ basis it would be inconvenient to pick up suitable values of TM transformation angles to get the three Jacobi momenta for the c.m. and intrinsic motion. So when we focus on the systems consisting of more than two constituents, it would be most practical to choose the $x$-dependent $b$ basis.

Here, we find that the $x$-dependent $b$ basis also converges better than the $x$-independent $b$ basis. To make comparisons with the $x$-independent $b$ basis, we calculate the two non-divergent form factors $F_2(t)$ and $B(t)$ with both bases at $N_\text{max} = K_\text{tot} - 1/2 = 80$, and compare them with the perturbative results with cutoffs associated with fixed and x-dependent b, and without cutoff at all.

\begin{figure*}[t!]
    \includegraphics[width=0.49\textwidth]{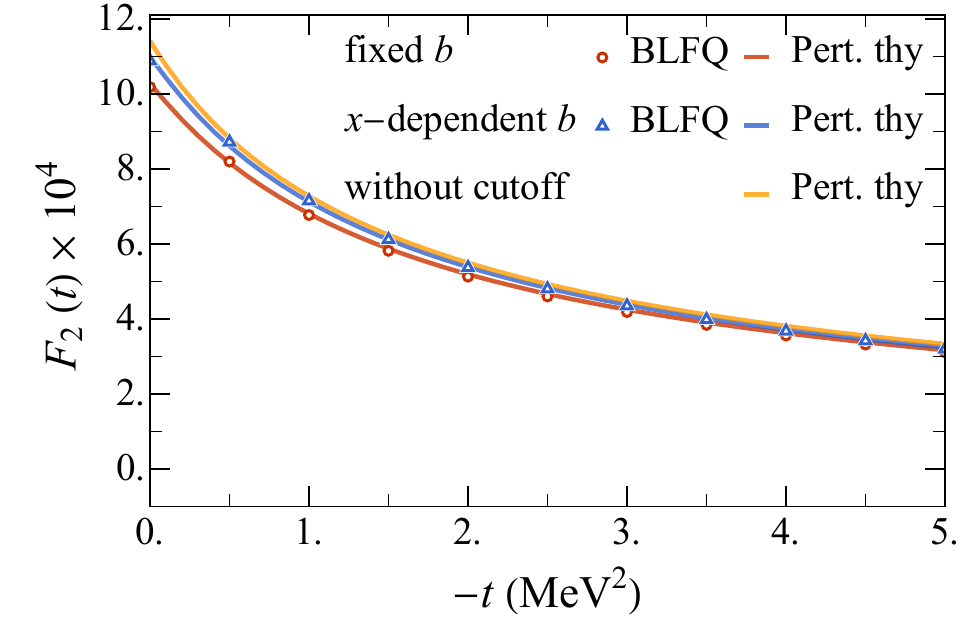}
    \includegraphics[width=0.49\textwidth]{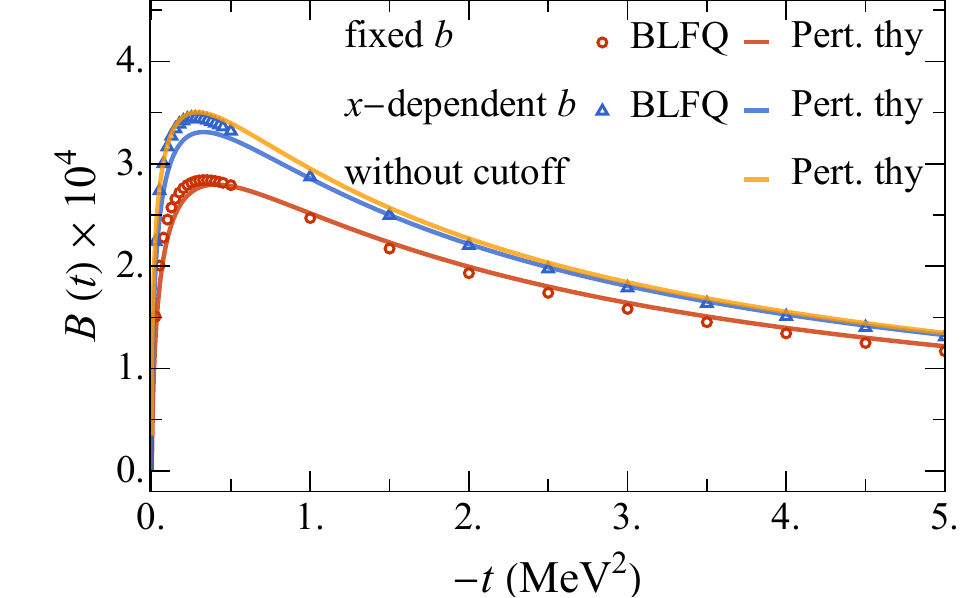}	
       %\hspace{-1.5cm}
	%	\includegraphics[scale=0.53]{F2com.pdf}
	%\end{subfigure}
	%\hspace{1cm}
	%\begin{subfigure}[t]{0.45\textwidth}
	%	\includegraphics[scale=0.53]{FFBcom.pdf}
	%\end{subfigure}
	\caption{\justifying{Two non-divergent form factors, the Pauli form factor $F_2$ and the gravitational form factor $B$, calculated with $x$-dependent $b$ and fixed-$b$ basis in BLFQ at $N_\text{max} = K_\text{tot}-1/2 = 80$, and compared with perturbation theory. The yellow curves label the perturbative results at infinite $N_\text{max}$. The $x$-dependent $b$ basis shows a faster convergence compared to the fixed-$b$ basis in the lower $-t$ region.}}
	\label{fig_F2B}
\end{figure*}

At infinite $N_\text{max}$ and $K_\text{tot}$, the BLFQ results with both bases are expected to converge to the perturbative results without cutoff. In the truncated bases, one can see from FIG.~\ref{fig_F2B} that for both $F_2(t)$ and $B(t)$, the $x$-dependent $b$ basis leads to values much closer to the perturbation theory than the fixed-$b$ basis. Meanwhile, one should be aware that the advantage of the $x$-dependent $b$ basis is more pronounced in the low $-t$ region. When $-t$ increases, the difference among three methods diminishes. Thus the $x$-dependent $b$ basis is more suitable to be used to calculate the observables associated with lower $-t$, such as the charge and mass radius.

\begin{figure}
	\centering
	\includegraphics[width=0.49\textwidth]{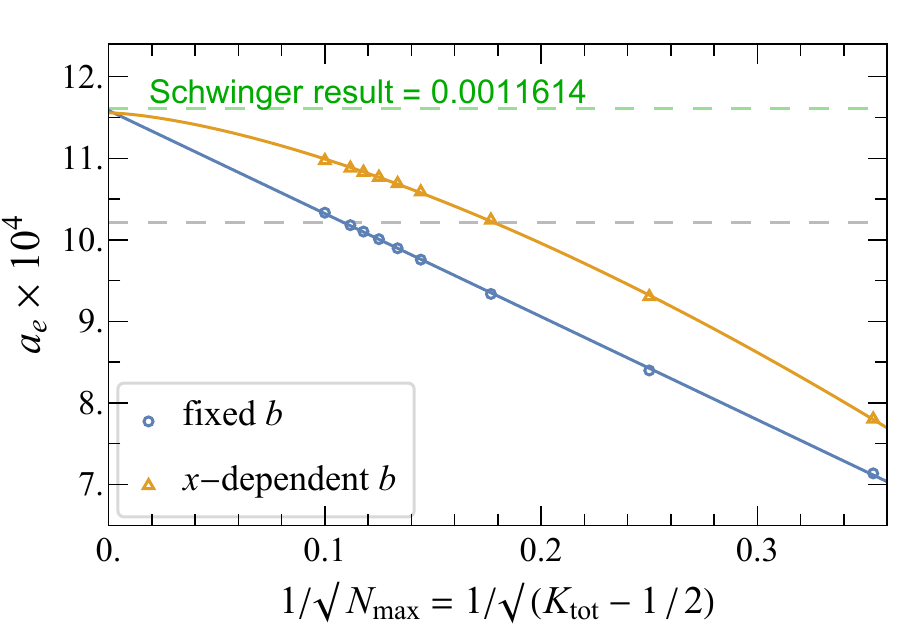}
	\caption{\label{fig_convg}\justifying{The convergence of two bases as $N_\text{max} $ approaches infinity. A faster convergence of $a_e \equiv F_2(t \rightarrow 0)$ is achieved with the $x$-dependent $b$ basis. For instance, to achieve $a_e \approx 0.00102$  (gray dashed line), which is about $88\%$ of the Schwinger value $\alpha^2/(2\pi)$, one needs $N_\text{max} =80 $ (with the basis dimension 5182401) in the fixed-$b$ basis, whereas one needs only $N_\text{max} = 32$ (with the basis dimension 134913) in the $x$-dependent $b$ basis.
		We apply linear extrapolation for the fixed-$b$ results, and the function $k (1/\sqrt{N_\text{max}})^{3/2}+m$ to fit the $x$-dependent $b$ results. The extrapolation of the results performed in both bases to infinite $N_\text{max}=K_\text{tot}-1/2$ agrees reasonably well with the Schwinger result.}}
\end{figure}

To further present the advantage of $x$-dependent $b$ basis for improving convergence, we compare the results of the electron anomalous magnetic moment $a_e$ calculated with two bases at different $N_\text{max}$ in FIG.~\ref{fig_convg}. We notice that comparable values can be achieved with the $x$-dependent $b$ basis at a much smaller $N_\text{max}$ compared to the fixed-$b$ basis, which translates to a much smaller basis dimensionality and thus greatly reduced computational resources. Again, in the infinite basis limit, both methods provide the results consistent with the Schwinger result. 

\subsection{Mass and Wave Function Renormalization}

\begin{figure}
	\centering
	\includegraphics[width=0.49\textwidth]{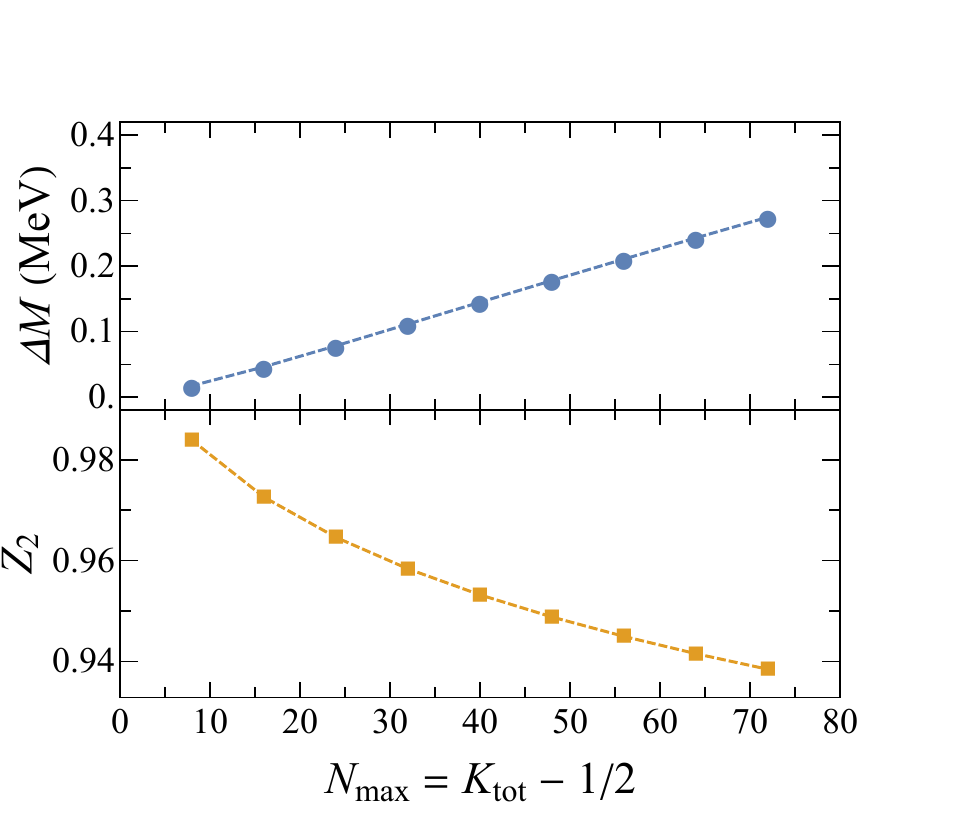}
	\caption{\label{fig_renorm}\justifying{Upper and lower panels are the electron mass counter-term $\Delta M$ and the wave function renormalization $Z_2$ as functions of $N_\text{max}$, respectively. The constant $b'$ in the basis scale parameter is chosen to be the physical electron mass, namely, $b' =M_e =0.511$ MeV.}}
\end{figure}

It should be noted that we need to perform two renormalization procedures in the diagonalization. First, we perform mass renormalization. In our current Fock sector truncation, due to the absence of the $\ket{ee\bar{e}}$ sector, a bare photon cannot fluctuate into an electron-positron pair.
Thus one only needs to consider the electron mass renormalization. According to the Fock sector-dependent renormalization approach~\cite{PhysRevD.77.085028,PhysRevD.86.085006}, when numerically diagonalizing the Hamiltonian matrix, we adjust the bare electron mass $m_e$ only in the $\ket{e}$ sector iteratively so that the resulting ground state mass of state $\ket{e_\text{phys}}$ is $M_e =0.511$ MeV. 
We introduce the mass counter-term  $\Delta M $ to denote the difference between physical electron mass and bare mass, that is, $\Delta M = m_e-M_e$. It compensates for the mass correction due to the quantum fluctuations to higher Fock sectors, namely the basis states in $\ket{e}$ sector coupled to those in $\ket{e\gamma}$ sector, which generates the conventional one-loop self-energy correction.
The dependence of the mass counter-term $\Delta M$ on the basis truncation parameters is shown in the upper panel of FIG.~\ref{fig_renorm}, and we can see that $\Delta M$ increases approximately linearly with increasing basis truncation parameters, which corresponds to the light-front perturbative results~\cite{BRODSKY2004333}.

Second, we perform wave function renormalization. In BLFQ, the Ward identity $Z_1 =Z_2$  is no longer held due to the  Fock sector truncation~\cite{BRODSKY2004333}. Here $Z_1$ is the renormalization factor for the vertex that couples the $\ket{e}$ and $\ket{e\gamma}$ sectors, which remains unity in the infinite basis limit with our Fock space truncation. $Z_2$ is the electron wave function renormalization, which can be interpreted as the probability of finding a bare electron in a physical electron system in the light-front dynamics, i.e., $Z_2 = \abs{\braket{e}{e_\text{phys}}}^2$. 
In order to remedy the artifacts from the violation of the Ward identity caused by the Fock sector truncation, following the previous works~\cite{PhysRevD.89.116004,ZHAO201465}, we rescale our naive results according to $Z_2$. Specifically, the rescaled observable is
\begin{equation}
\expval{\mathcal{O}^\text{re}}=\frac{2Z_2-1}{Z^{2}_2}  \expval{ P^{-1}_e \mathcal{O} P_e} +\frac{1}{Z_2}\expval{ P^{-1}_{e\gamma} \mathcal{O} P_{e\gamma}}.
\label{eq_renorm}
\end{equation}
Here $\mathcal{O}$ denotes the operators for the form factors and the GPDs. $P_e$ and $P_{e\gamma}$ are the projection operators onto the $\ket{e}$ and $\ket{e\gamma}$ sectors, respectively. The dependence of the wave function renormalization $Z_2$ on the basis truncation parameters is shown in the lower panel of FIG.~\ref{fig_renorm}.
$Z_2$ tends to be zero in the infinite basis limit.

\section{\label{sec4}Numerical Results}

We perform our calculations in truncated bases with truncation parameters $K_\text{tot}$ and $N_\text{max}$, where $K_\text{tot}$ represents the longitudinal resolution, and $N_\text{max}$ specifies the ultraviolet (UV) and inferred (IR) regulators in the transverse plane. For simplicity, we set $N_\text{max} = K_\text{tot} - 1/2$ throughout this paper.
We set the coupling strength $\alpha=1/137.036$ and the constant $b'$ in the scale parameter $b$ of the transverse basis function to be the physical electron mass, according to Ref.~\cite{ZHAO201465}, i.e. $b' =M_e =0.511$ MeV, and $b_i =b' \sqrt{x_i} =M_e \sqrt{x_i}$. We then iteratively alter the mass counter-term $\Delta M$ to fit the physical mass, i.e. $M=M_e$, and we employ the LFWFs to calculate the electromagnetic form factors, the gravitational form factors and the GPDs for the electron.

We also calculate the light-front perturbation results to compare with our BLFQ results.
For the perturbative calculation, we use the wave functions in Ref.~\cite{BRODSKY200199,BRODSKY2001311}. For the two-particle Fock state, we have
\begin{equation}
\left\{\begin{array}{l}
\psi_{+\frac{1}{2}+1}^{\uparrow}\left(x, \vec{k}_{\perp}\right)=-\sqrt{2} \frac{\left(-k^1+\mathrm{i} k^2\right)}{x(1-x)} \varphi, \\
\psi_{+\frac{1}{2}-1}^{\uparrow}\left(x, \vec{k}_{\perp}\right)=-\sqrt{2} \frac{\left(+k^1+\mathrm{i} k^2\right)}{1-x} \varphi, \\
\psi_{-\frac{1}{2}+1}^{\uparrow}\left(x, \vec{k}_{\perp}\right)=-\sqrt{2}\left(M-\frac{m}{x}\right) \varphi, \\
\psi_{-\frac{1}{2}-1}^{\uparrow}\left(x, \vec{k}_{\perp}\right)=0,
\end{array}\right.
\end{equation}
and
\begin{equation}
\left\{\begin{array}{l}
\psi_{+\frac{1}{2}+1}^{\downarrow}\left(x, \vec{k}_{\perp}\right)=0, \\
\psi_{+\frac{1}{2}-1}^{\downarrow}\left(x, \vec{k}_{\perp}\right)=-\sqrt{2}\left(M-\frac{m}{x}\right) \varphi, \\
\psi_{-\frac{1}{2}+1}^{\downarrow}\left(x, \vec{k}_{\perp}\right)=-\sqrt{2} \frac{\left(-k^1+\mathrm{i} k^2\right)}{1-x} \varphi, \\
\psi_{-\frac{1}{2}-1}^{\downarrow}\left(x, \vec{k}_{\perp}\right)=-\sqrt{2} \frac{\left(+k^1+\mathrm{i} k^2\right)}{x(1-x)} \varphi ,
\end{array}\right.
\end{equation}
for the parallel and anti-parallel helicity states, respectively, where
\begin{equation}
\begin{aligned}
\varphi=&\varphi\left(x, \vec{k}_{\perp}\right)\\
=&\frac{e / \sqrt{1-x}}{M^2-\left(\vec{k}_{\perp}^2+m^2\right) / x-\left(\vec{k}_{\perp}^2+\lambda^2\right) /(1-x)} ,
\end{aligned}
\end{equation}
$M$ is the physical electron mass, $m$ is the bare electron mass, $\lambda$ is the photon mass, and $x$ and $\vec{k}_\perp$ are the longitudinal momentum fraction and the transverse momentum of the constituent electron, respectively. Here we take $M=m$ and keep a small nonzero value of the photon mass.

In BLFQ, the truncation up to the one-fermion one-gauge-boson Fock state component is expected to contain the equivalent physics as the Schwinger one-loop radiative correction.  For the convergent observables, the leading-order perturbation theory is expected to give near-identical results with BLFQ at infinite basis size in the case of the physical electron. For finite basis, we need to take into account of the UV and IR cutoffs. In BLFQ, $N_\text{max}$ and $K_\text{tot}$ work as the underlying regulators; while for perturbation theory, for purposes of comparison, we manually implement the corresponding integral cutoffs in  momentum space. In this work, $x\equiv p^+_e/P^+= p^+_e/(p^+_e+p^+_\gamma)$ stands for the longitudinal momentum fraction of the constituent electron. In the transverse directions, since the scale parameter is associated with $x_i$, the UV and IR cutoffs are chosen to match those in BLFQ which are estimated as $\Lambda_{\text{UV}}  \approx b'\sqrt{2x(1-x)N_\text{max} }$ and $\lambda_{\text{IR}} \approx b'\sqrt{x(1-x)/(2N_\text{max})}$~\cite{PhysRevD.88.065014,PhysRevD.89.116004}. In the longitudinal direction, we integrate from $x$=$0.5/K_\text{tot}$ to $1-1/K_\text{tot}$.
For the contribution of the bare electron Fock state, we adopt the corresponding renormalization, see Ref.~\cite{BRODSKY2001311}.

\begin{figure*}
    \includegraphics[width=0.49\textwidth]{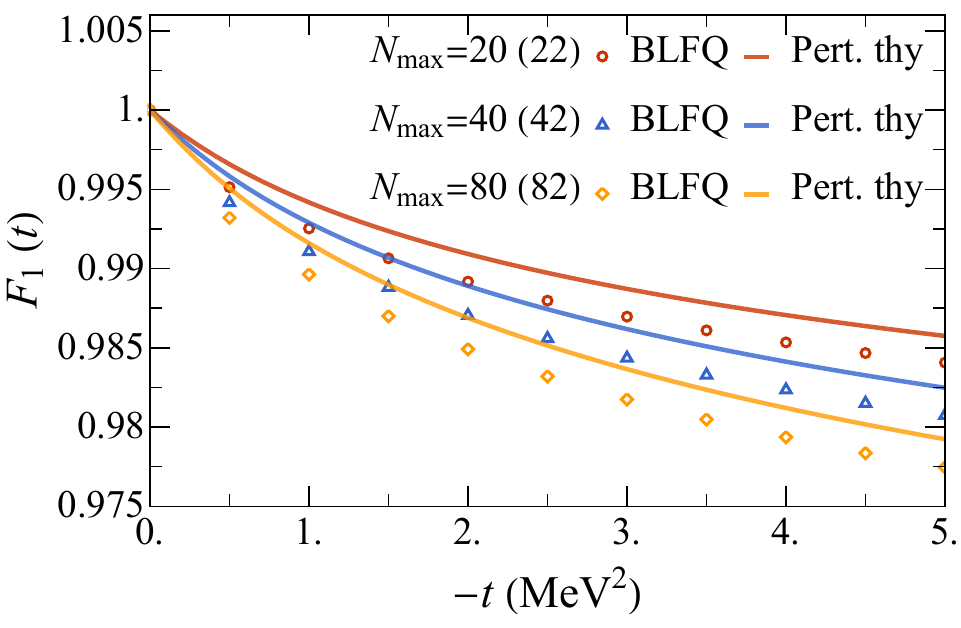}
    \includegraphics[width=0.49\textwidth]{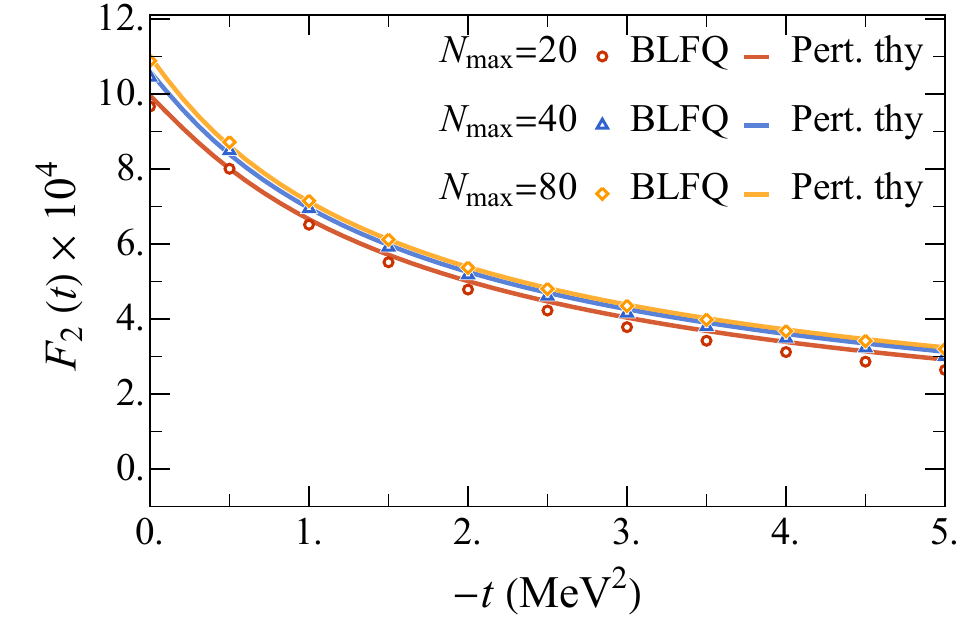}
	%\hspace{-1cm}
	%\begin{subfigure}[t]{0.5\textwidth}
	%	\centering
	%	\includegraphics[scale=0.5]{F1de.pdf} 
	%\end{subfigure}
	%\hspace{.5cm}
	%\begin{subfigure}[t]{0.5\textwidth}
	%	\centering
	%	\includegraphics[scale=0.5]{F2de.pdf}
	%\end{subfigure}
	\caption{\label{fig_eff}\justifying{The electromagnetic form factors $F_1$ and $F_2$ calculated with BLFQ and with light-front perturbation theory. The BLFQ results of $F_1$ are given as the average of adjacent $N_\text{max}$, e.g. $N_\text{max}=20\ (22)$ denotes that the results are the average of $N_\text{max}=20$ and $N_\text{max}=22$, to smooth over the ``odd-even'' effect, see text. For $F_2$, we only adopt the results with even $N_\text{max}/2$ for faster convergence. }}
\end{figure*}

\begin{figure}
    \centering
    \includegraphics[width=0.49\textwidth]{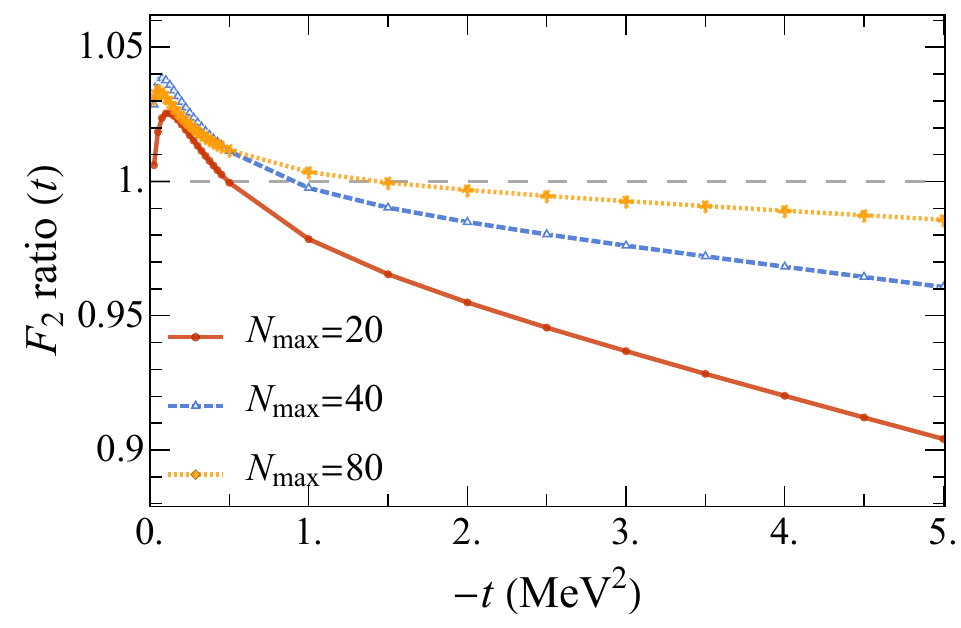}
    \caption{\label{fig_f2r}\justifying{The ratio of the BLFQ results to the perturbative results for $F_2$. The dots are the results of the ratio for different $-t$ and different $N_\text{max}$, and we connect them to make the trend more clear. With larger $N_\text{max}$, the ratio becomes closer to 1 (gray dashed line).}}
\end{figure}

\begin{figure*}
    \includegraphics[width=0.49\textwidth]{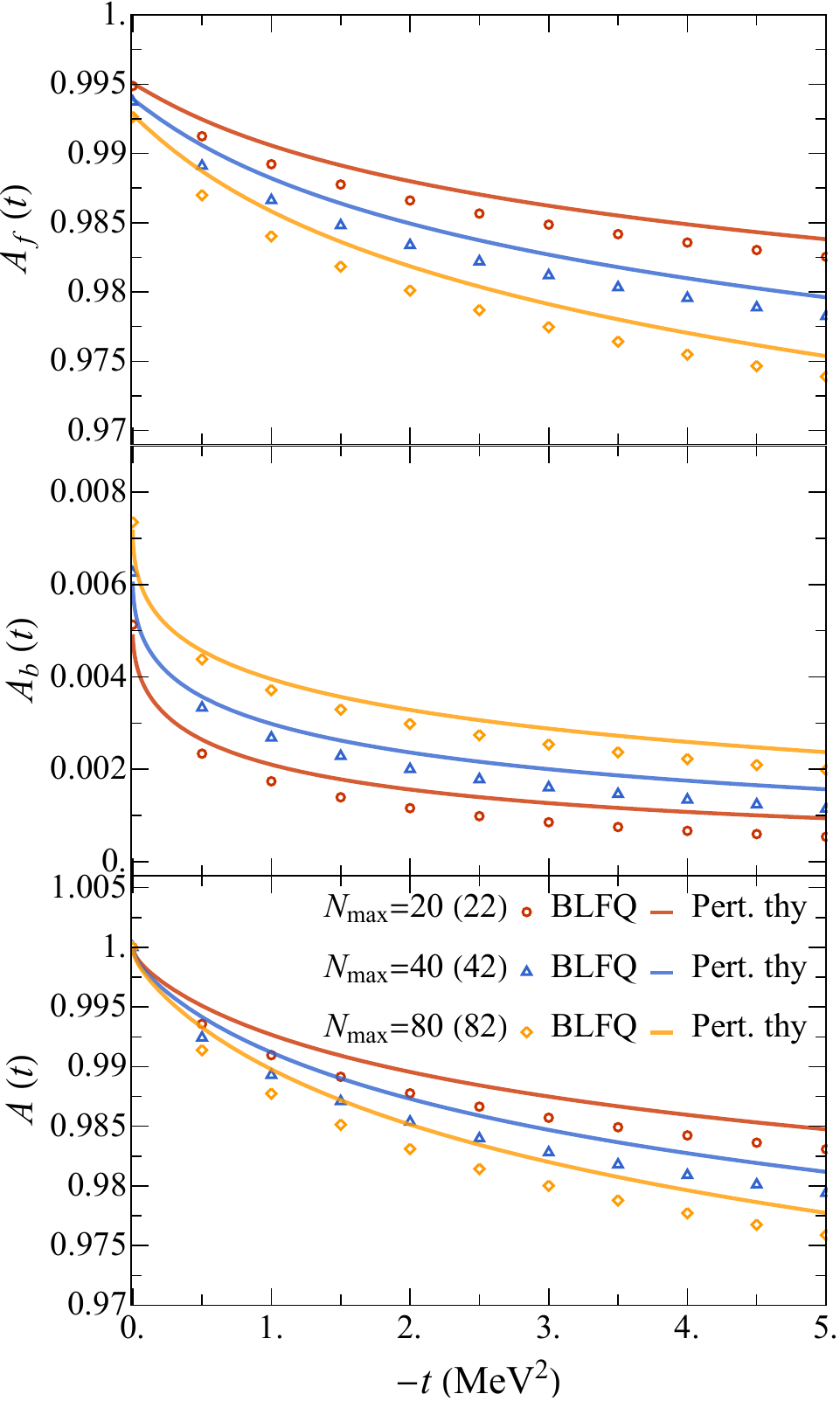}
    \includegraphics[width=0.49\textwidth]{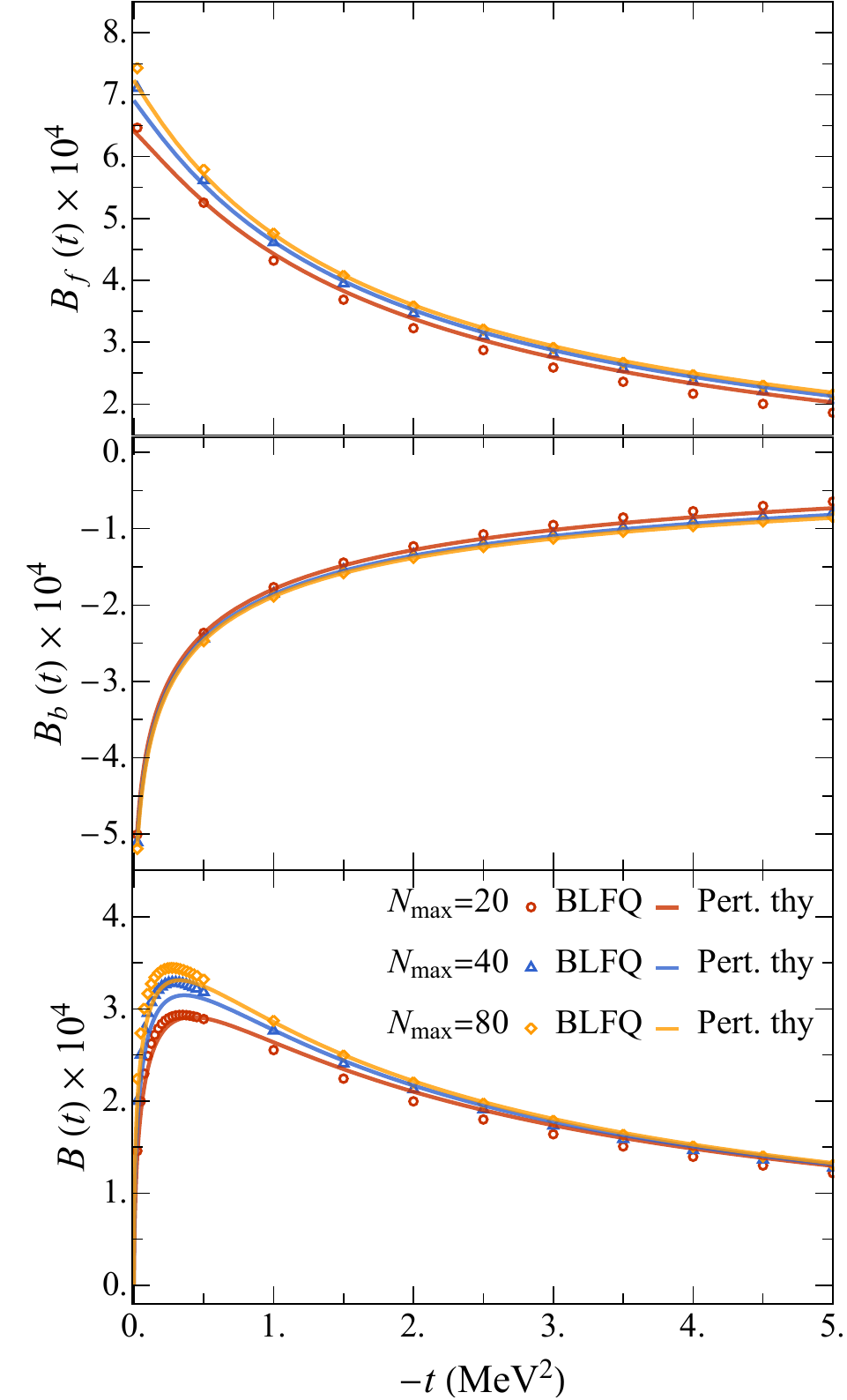}
	%\centering
	%\begin{subfigure}[t]{0.5\textwidth}
	%	\centering
	%\end{subfigure}
	%\hspace{.5cm}
	%\begin{subfigure}[t]{0.5\textwidth}
	%	\centering
	%	\includegraphics[scale=0.5]{GFFBde.pdf}
	%\end{subfigure}
	\caption{\label{fig_gff}\justifying{We show the gravitational form factor contributions from the constituent electron and photon respectively, then followed by the total gravitational form factors. Remarkably, $B(t \rightarrow 0)$ from BLFQ is always zero independent of $N_\text{max}$. Similar to $F_2$, for the convergent observables $B_\text{f,b}(t)$ and their summation $B(t)$, the discrepancy between BLFQ and light-front perturbation theory decreases as the basis size increases.}}
\end{figure*}

FIG.~\ref{fig_eff} illustrates the electromagnetic form factors as functions of $-t$ for different basis truncations. We use open markers for the BLFQ results, and compare them with the light-front perturbation theory for the two-particle Fock state of the electron (solid curves)~\cite{BRODSKY2001311}. 
 
Since the UV and IR cutoffs in BLFQ and those in perturbation theory are implemented in different manners, we cannot expect the results from these two methods to exactly match. Nevertheless, for the convergent observable $F_2$, the differences between the two methods are expected to decrease with the increase of $N_\text{max} = K_\text{tot} -1/2$, as we show in FIG.~\ref{fig_eff}. To study this convergence in more detail,, we show the ratio of the BLFQ results to the perturbative results for $F_2$ in FIG.~\ref{fig_f2r}. Of course, one does not expect the perturbative and non-perturbative results to become exactly equal but one does anticipate that the fully  convergent ratio should be indistinguishable from unity on the scale of FIG.~\ref{fig_f2r}. With larger $N_\text{max}$, the ratio becomes closer to 1, which is more obvious in the large $-t$ region. In the low $-t$ region, the ratio first goes further away from 1 at small $N_\text{max}$ and then goes back as $N_\text{max}$ keeps increasing.  

We also observe an “odd-even effect” in BLFQ where the results fall into two groups with even and odd $N_\text{max}/2$ (see Ref.~\cite{ZHAO201465}). This effect is due to the oscillatory behavior of the transverse basis functions. Thus for $F_1(t)$, we average over results obtained at two adjacent $N_\text{max}$ values to smooth this odd-even effect. For the corresponding perturbation theory results, we adopt the averaged $N_\text{max}$ value to match the UV and IR cutoffs. For example, we adopt $N_\text{max} = 21$ in the UV and IR cutoffs to match $N_\text{max} = 20(22)$ in BLFQ. See FIG.~\ref{fig_eff}.
Unlike the Dirac form factor, the Pauli form factor is convergent in the entire momentum space. However, we still apply the cutoffs to the perturbation theory to keep consistency with BLFQ. Without taking the average between adjacent $N_\text{max}$, we obtain reasonable agreement between two approaches. 
In particular, $F_2(t \rightarrow 0)$ corresponds to the anomalous magnetic moment $a_e$. In our previous work~\cite{ZHAO201465}, upon extrapolating $N_\text{max}$ and $K_\text{tot}$ to infinity, the resulting $a_e$ agrees with Schwinger result to an accuracy of $0.06\%$, which is consistent with the expected numerical precision. 
Both electromagnetic form factors decrease as the momentum transfer increases. $F_2$ will fall to zero at infinite $-t$, while $F_1$ will end up with $(2Z_2-1)/Z_2$ (c.f. Eq.~\eqref{eq_renorm}), the contribution from the single electron sector which is independent on $-t$~\cite{PhysRev.126.2256}.

Similar to the electromagnetic form factors, we adopt the averaging method for the gravitational form factor $A(t)$ to smooth the influence from the odd-even effect, and we adopt even $N_\text{max}/2$ for the gravitational form factor $B(t)$. The evaluation of gravitational form factors is similar to that for electromagnetic form factors albeit with some additional complexity: both the constituent fermion and boson couple to the graviton and thus contribute to the gravitational form factors. The matrix elements of  $T^{++}$ allow us to calculate the fermion and boson contributions to the gravitational form factors $A_\text{f,b}(t)$ and $B_\text{f,b}(t)$ separately.

According to Ji sum rule~\cite{PhysRevLett.78.610,PhysRevD.58.056003}, we have 
\begin{equation}
\begin{aligned}
A(0)=A_\text{f}(0)+A_\text{b}(0)&=(P_\text{f}+P_\text{b})/P_\text{tot}=1,\\
\frac{1}{2}\big[A_\text{f}(0)+B_\text{f}(0)+A_\text{b}(0)&+B_\text{b}(0)\big]=J_\text{f}+J_\text{b}=\frac{1}{2},
\end{aligned}
\end{equation}
at zero momentum transfer. These are consequences of the conservations of momentum and angular momentum, and lead to $B(0)=B_\text{f}(0)+B_\text{b}(0)=0$. This last result is known as the vanishing anomalous gravito-magnetic moment which is closely connected with the Einstein equivalence principle~\cite{Teryaev_2007,Teryaev2016}. 
In BLFQ we obtain these results at all basis sizes as shown in FIG.~\ref{fig_gff}. Specially, $B_\text{b} (t)$ is negative when $-t>0$ and thus consistent with the corresponding GPD $E_\text{b}$, which will be discussed later. 

\begin{figure*}
    \includegraphics[width=0.49\textwidth]{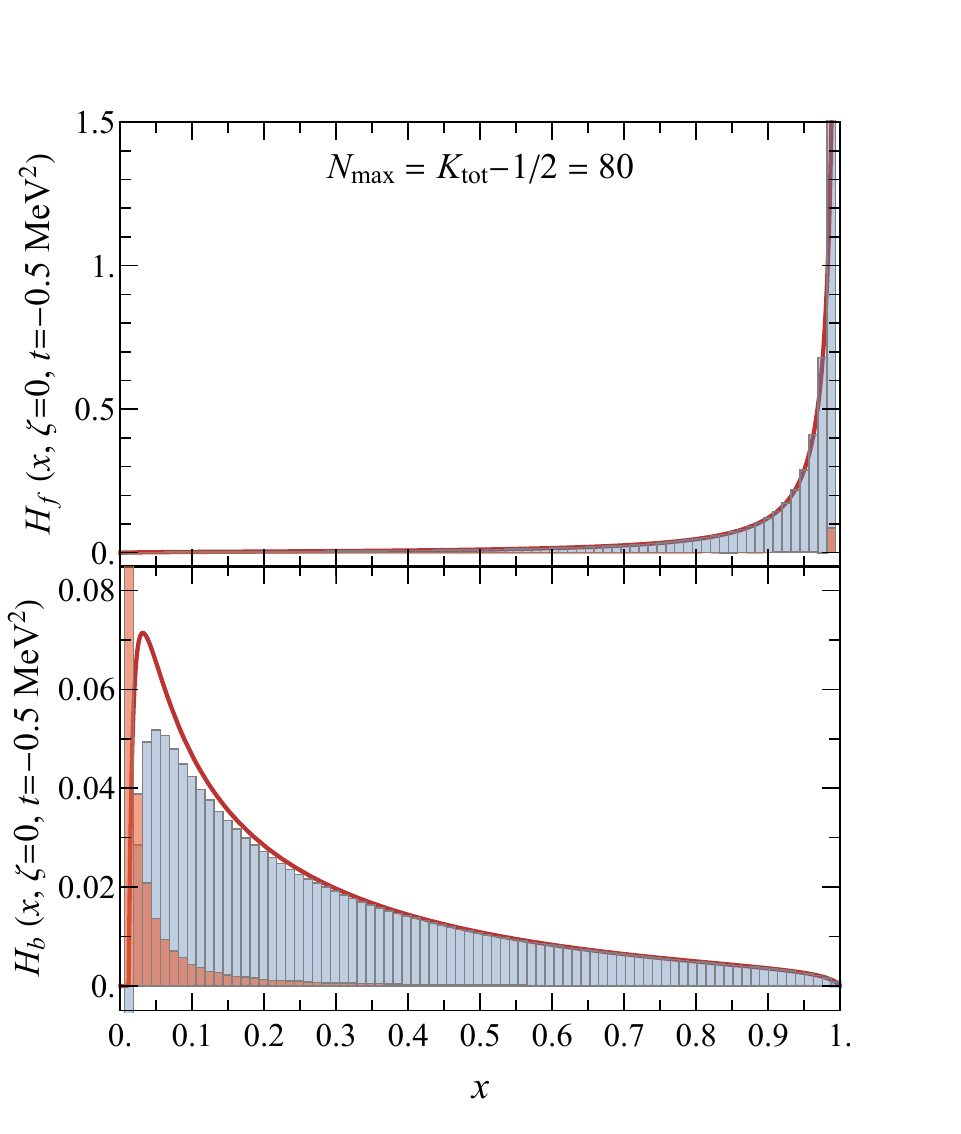}
    \includegraphics[width=0.49\textwidth]{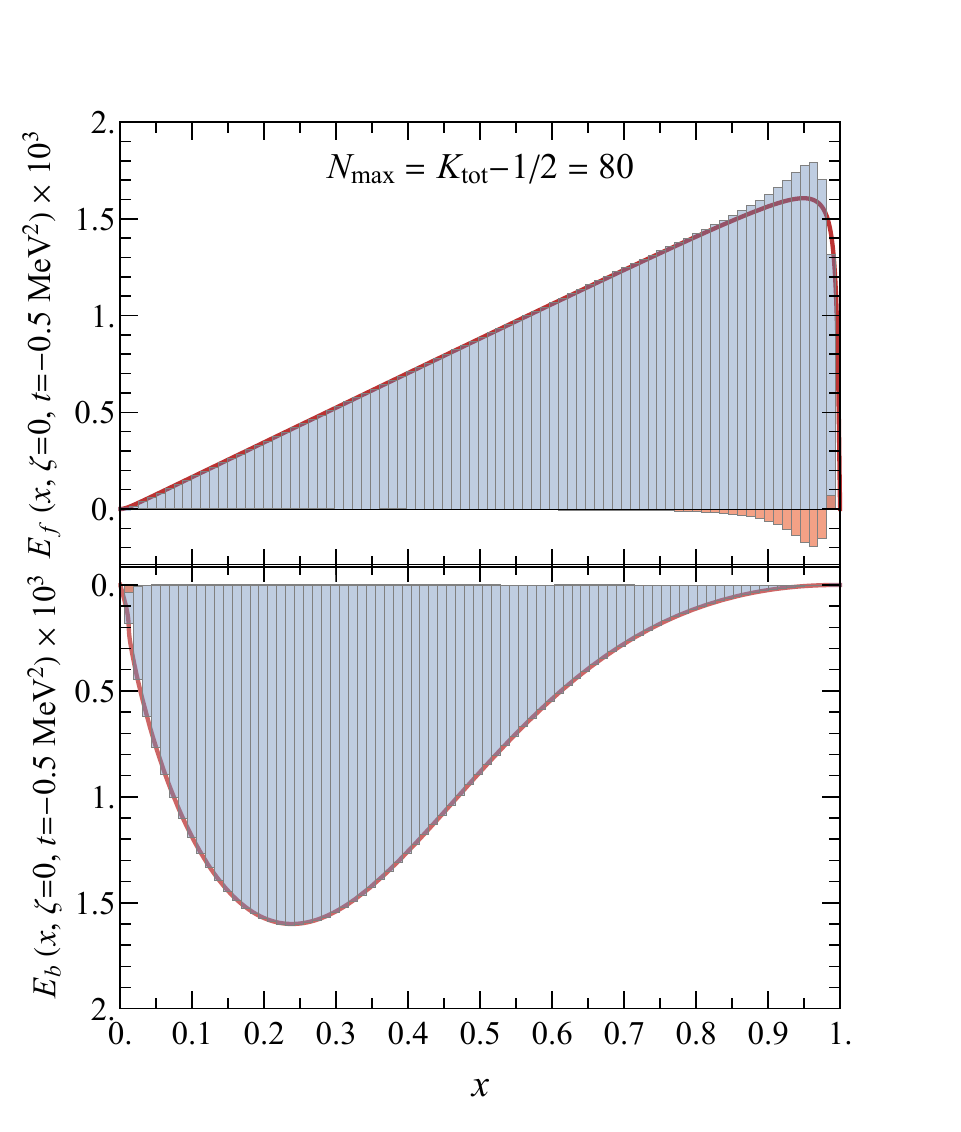}	
 %\hspace{-1cm}
	%\centering
	%\begin{subfigure}[t]{0.45\textwidth}
	%	\centering
	%	\includegraphics[scale=0.5]{t2H.pdf}
	%\end{subfigure}
	%\hspace{1cm}
	%\begin{subfigure}[t]{0.45\textwidth}
	%	\centering
	%	\includegraphics[scale=0.5]{t2E.pdf}
	%\end{subfigure}
    \caption{\label{fig_gpd2D}\justifying{The GPDs of electron and photon in a physical electron, labeled as ``$H_\text{f} \ (E_\text{f})$'' and ``$H_\text{b} \ (E_\text{b})$'', respectively. We present the GPDs calculated at $t=-0.5 \text{ MeV}^2$ and $N_\text{max} = K_\text{tot}-1/2=80$. Blue bars show the value obtained by BLFQ at $(0.5+i)/K_\text{tot} \ (i=0,1,2,...,79)$ for $H_\text{f} \ (E_\text{f})$ and at $1-(0.5+i)/K_\text{tot}$ for $H_\text{b} \ (E_\text{b})$ respectively, and are compared with the light-front perturbation theory results, which are shown with red solid curves. We also use the red bars to present the difference between BLFQ and the perturbation theory from $i/K_\text{tot}$ to $(i+1)/K_\text{tot}$ for $H_\text{f} \ (E_\text{f})$ and from $1-(i+1)/K_\text{tot}$ to $1-i/K_\text{tot}$ for $H_\text{b} \ (E_\text{b})$, respectively. In the plot of GPD $H_\text{b}$, the red curve (light-front perturbation theory results) dives fast to nearly zero when $x$ reaches $0$, which results from the influence of the UV cutoff we set, see text.}}
\end{figure*}

\begin{figure*}
    \includegraphics[width=0.49\textwidth]{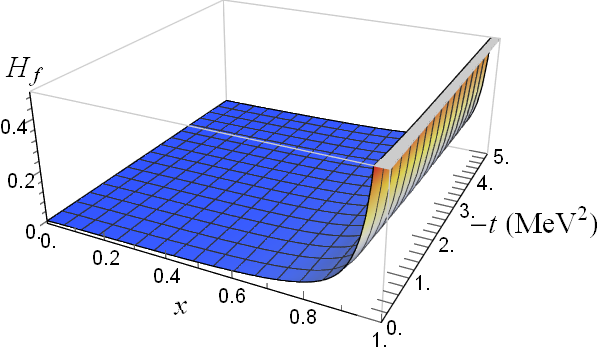}
    \includegraphics[width=0.49\textwidth]{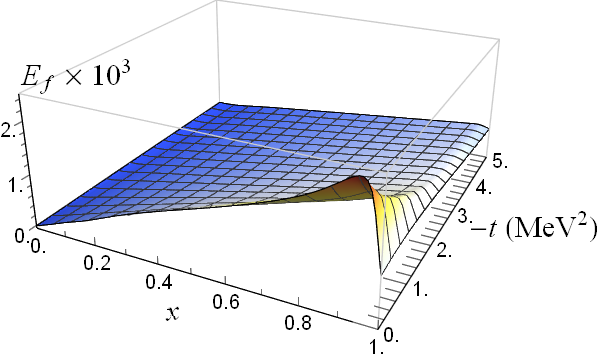}
    \\
    \includegraphics[width=0.49\textwidth]{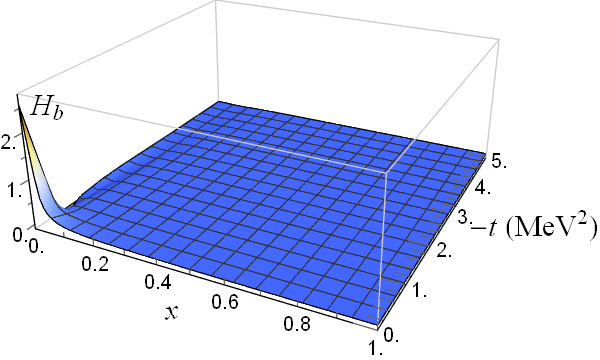}
    \includegraphics[width=0.49\textwidth]{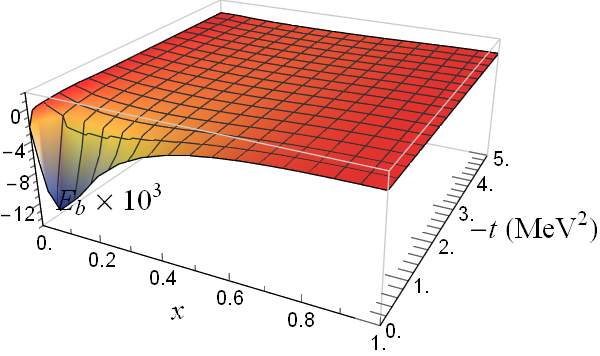}
	\caption{\label{fig_gpd3D} \justifying{3D plots of GPDs of constituent fermion and boson with respect to $x$ and $-t$ calculated in BLFQ at basis $N_\text{max}=K_\text{tot}-1/2=80$.}}
\end{figure*}

In order to further our understanding of the differences between results from BLFQ and light-front perturbation theory, we turn to the unpolarized GPDs, $H$ and $E$ of the physical electron.
It has been shown in a previous work calculated with the fixed $b$ basis that GPDs $H(x,\zeta=0,t)$ and $E(x,\zeta=0,t)$ agree reasonably well with light-front perturbation theory~\cite{PhysRevD.89.116004}, while our work is with the $x$-dependent $b$ basis. We take $x$ as the longitudinal momentum fraction of the consitituent parton for the results of GPDs.
As an example, we present the results of GPDs in FIG.~\ref{fig_gpd2D} and compare them with perturbation theory at $t = -0.5\text{ MeV}^2$.
$H_\text{f}$ increases with $x$. $H_\text{b}$ grows rapidly from nearly zero to around $0.07$ in the $x \sim 0$ region which results from the influence of the UV cutoff, and then decreases as $x$ increases. $E_\text{f}$ first increases as $x$ increases and drops quickly when $x$ reaches 1 due to the reducing UV cutoff at large $x$. $E_\text{b}$ is negative and has a peak near $x=0.2$.
We can see that the main difference between BLFQ and light-front perturbation theory is in the $x\rightarrow 1$ region for $H_\text{f} \ (E_\text{f})$ and in the $x\rightarrow 0$ region for $H_\text{b} \ (E_\text{b})$, respectively, where the constituent photon takes a small longitudinal momentum fraction. Due to the divergent behavior of $H$ at these regions, the results are sensitive to the details of the cutoff. Hence, the differences between the two light-front approaches are larger in $F_1(t)$ and $A(t)$ than those in $F_2(t)$ and $B(t)$.
Especially, since the cutoffs are associated with the factor $\sqrt{x(1-x)}$, the difference between the UV cutoff and the IR cutoff decreases rapidly as $x$ approaches 0, which makes the $H_ \text{b}$ results of the perturbation theory dive to nearly zero in the small $x$ region.
%However, in all cases, the differences between the results of the convergent observables are decreasing systematically as the cutoffs are lifted, indicating the utility of our adopted relationship of the cutoffs between the two different methods. 

In order to provide a visual overview of the general structure in both the transverse and longitudinal direction, we show the GPDs as 3-dimensional (3D) plots as functions of both $x$ and $t$ in FIG.~\ref{fig_gpd3D}. The GPD $H$ of the constituent electron is divergent at large $x$ region no matter what the value of $t$ is, while the GPD $H$ of the constituent photon diverges at small $x$ only at $t$ around 0. The GPD $E$ of the constituent electron has its peak at the large $x$ region, while the GPD $E$ of the constituent photon being negative also has a peak.

\section{Summary}
\label{sec5}
In this work, we investigated the physical electron system in the BLFQ approach, where two leading Fock sectors were considered. Based on the resulting LFWFs we calculated both the electromagnetic and the gravitational form factors and their corresponding GPDs.
We performed nonperturbative renormalization both on the light-front Hamiltonian and on the resulting LFWFs.
All these results show reasonable agreement with light-front perturbation theory calculated with regulators that correspond to the truncated basis representation of the LFWFs. Especially, for the convergent observables $F_2(t)$ and $B(t)$, the agreement between the perturbative and BLFQ results improves as the basis size increases. 
We also compared the results using the $x$-dependent $b$ basis with those using the $x$-independent $b$ basis. We demonstrated that the $x$-dependent $b$ basis results in faster convergence to the anticipated perturbative results for the electromagnetic and gravitational form factors.
These results not only validate the BLFQ approach but also provide guidance for efficient implementation of computational approaches to more complex QED and QCD systems in the light-front Hamiltonian framework.

The agreement with perturbation theory on the form factors of the electron constitutes a comprehensive test of the LFWFs obtained from the BLFQ approach. It is also an important test of the viability of the nonperturbative renormalization procedure carried out in BLFQ. This application of BLFQ to the physical electron system provides us with the guidance to study the bound states in QCD with BLFQ where Fock sectors beyond the valence sector are included in the basis. As a next step, we plan to investigate the gravitational form factors of hadrons, for instance the proton~\cite{rajinprepara}, compare the results with those from existing experiments~\cite{PhysRevD.97.014020} and other theoretical approaches, and make predictions for future experiments. 
Another line of future development is to include even higher Fock sectors, eg. $\ket{ee\bar{e}}$, in the basis to further test the approach. By doing so, we need to further develop the nonperturbative renormalization procedure to handle, for example, the possible divergences arising from the quantum fluctuation of a photon to an electron-positron pair. If this is successful, we will be able to obtain a more realistic description of the relativistic bound state structure in the BLFQ approach.

\section{Acknowledgments}			
We appreciate the technical assistance from Pieter Maris, and also wish to thank M. Li, W. Qian, A. Yu, S. Xu, S. Nair, J. Wu, Z. Zhu, Y. Liu and C. Mondal for fruitful discussions.
This work was supported in part by the U.S. Department of Energy under Grant No. DE-SC0023692, and in part by the Chinese Academy of Sciences under Grant No. YSBR-101. This research used resources of the National Energy Research Scientific Computing Center (NERSC), a U.S. Department of Energy Office of Science User Facility located at Lawrence Berkeley National Laboratory, operated under Contract No. DE-AC02-05CH11231 using NERSC award NP-ERCAP0020944. Guo-Li Wang is supported by the National Natural Science Foundation of China (NSFC) under the Grants No. 12075073. Y. L. is supported by the New faculty start-up fund of the University of Science and Technology of China. X. Z. is supported by new faculty startup funding by the Institute of Modern Physics, Chinese Academy of Sciences, by Key Research Program of Frontier Sciences, Chinese Academy of Sciences, Grant No. ZDBS-LY-7020, by the Natural Science Foundation of Gansu Province, China, Grant No. 20JR10RA067, by the Foundation for Key Talents of Gansu Province, by the Central Funds Guiding the Local Science and Technology Development of Gansu Province, Grant No. 22ZY1QA006, by Gansu International Collaboration and Talents Recruitment Base of Particle Physics (2023-2027), by International Partnership Program of the Chinese Academy of Sciences, Grant No. 016GJHZ2022103FN, by National Natural Science Foundation of China, Grant No. 12375143 and by the Strategic Priority Research Program of the Chinese Academy of Sciences, Grant No. XDB34000000.

\bibliography{eff.bib}

%\end{CJK}	
\end{document}